\documentclass{emulateapj}

\shorttitle{The Column Density Variance-${\cal M}_s$ Relationship}

\shortauthors{BURKHART ET AL.}
\begin{document}

\title{The Column Density Variance-${\cal M}_s$ Relationship}

\author{Blakesley Burkhart\altaffilmark{1} and A. Lazarian\altaffilmark{1}}
\affil{$^1$ {Astronomy Department, University of Wisconsin, Madison, 475 N.  Charter St., WI 53706, USA}}

\begin{abstract}
Although there are a wealth of column density tracers for both the molecular and diffuse interstellar medium,
there are very few observational studies investigating the relationship between the density variance ($\sigma^2$) and the sonic Mach number (${\cal M}_s$).
This is in part due to the fact that the  $\sigma^2$-${\cal M}_s$ relationship
is derived, via MHD simulations, for the 3D density variance only, which is not a direct observable. 
We investigate the utility of a 2D column density $\sigma_{\Sigma/\Sigma_0}^2$-${\cal M}_s$ relationship using 
solenoidally driven isothermal MHD simulations and find that the best fit follows closely the 
form of the 3D density $\sigma_{\rho/\rho_0}^2$-${\cal M}_s$ trend
but includes a scaling parameter $A$ such that:
$\sigma_{ln(\Sigma/\Sigma_0)}^2=A\times ln(1+b^2{\cal M}_s^2)$, where $A=0.11$ and $b=1/3$.
This relation is consistent with the observational data reported for the Taurus 
and IC 5146 molecular clouds with $b=0.5$ and $A=0.16$ and $b=0.5$ and $A=0.12$, respectively.
These results open up the possibility of using the 2D column density values of $\sigma^2$ 
for investigations of the relation between the sonic Mach number and the PDF variance in addition to existing PDF sonic Mach number relations.
\end{abstract}

\section{Introduction}
\label{intro}

Magnetohydrodynamic (MHD) turbulence is a critical ingredient to include when considering the physics of the interstellar medium (ISM).
The turbulent nature of ISM gasses is evident from a variety of observations including
electron density fluctuations (see Armstrong
et al. 1994; Chepurnov \& Lazarian 2010), non-thermal broadening of emission and absorption lines, (e.g. CO,
HI, see Spitzer 1979; Stutzki \& Guesten 1990; Heyer \& Brunt 2004) and fractal and hierarchical structures in the diffuse and molecular ISM 
(see Elmegreen \& Elmegreen 1983; Vazequez-Semandeni 1994; Stutzki et al. 1998; Sanchez et al. 2005; Roman-Duval et al. 2010). 
A number of
new techniques, including those studying the turbulence velocity spectrum (see Lazarian 2009 for a review) and the sonic Mach number
 ${\cal M}_s \equiv \langle |{\bf v}|/C_s \rangle$ and Alfv\'en number ${\cal M}_A \equiv \langle |{\bf v}|/V_A \rangle$ of 
turbulence\footnote{The sound
and the Alf\'ven speed are denoted by $C_s$ and $V_A$, respectively.}  (see Kowal et al. 2007; Burkhart et al. 2009, 2010; Burkhart, Lazarian, \& Gaensler 2012; Esquivel \& Lazarian 2010; 
Tofflemire, Burkhart \& Lazarian 2011) have been developed recently.

An additional signature of a turbulent ISM is that the density and column density probability distribution functions (PDF) are expected to take a log-normal form (Vazquez-Semadeni 1994).\footnote{
The relation between turbulence and the log-normal distribution can be understood 
as a consequence of the multiplicative central limit theorem assuming that individual density perturbations are independent and random.}
Log-normal PDFs have been observed in multiple phases in the ISM including in molecular clouds (Brunt 2010), in dust extinction maps (Kainulainen et al. 2011; Kainulainen \& Tan 2012), 
and in the diffuse ISM (Hill et al. 2008; Berkhuijsen \& Fletcher 2008). 
Furthermore, the PDF was shown to be important for analytic models of star formation rates and initial mass functions  (Krumholz \& Mckee 2005;
Hennebelle \& Chabrier 2008,2011; Padoan \& Nordlund 2011).

The PDF of both the logarithmic and linear distribution of the gas
is an important method for determining ${\cal M}_s$. 
In general, the most common PDF-to-${\cal M}_s$ prescription
is to utilize numerical simulations to formulate an empirical relationship
between the PDF moments of the density or column density and relate these back to the sonic Mach number.
For example, several authors have suggested the turbulent sonic Mach number 
can be estimated from the calculation of the density variance (Padoan et al. 1997; Passot \& Vazquez-Semadeni 1998; Beetz et al. 2008; Price, Federrath, \& Brunt 2011, henceforth known as PFB11) 
and the column density skewness/kurtosis (Kowal et al. 2007; 
Burkhart et al. 2009, 2010). 
Other studies investigate the utility of finding ${\cal M}_s$ by fitting a function to the PDF and using the resulting fit parameters as 
descriptors of turbulence (e.g. the Tsallis function, Esquivel \& Lazarian 2010; Tofflemire, Burkhart, \& Lazarian 2011).

In particular, the relationship between  ${\cal M}_s$ and the variance of the logarithm of the density distribution 
as seen in numerical models (Padoan et al. 1997; Passot \& Vazquez-Semadeni 1998; PFB11) generally takes the form: 
\begin{equation}
 \sigma_{\rho/\rho_0}^2=b^2{\cal M}_s^2
\label{eq:linvar}
\end{equation}
where $\rho_0$ is the mean value of the 3D density field, $b$ is a constant of order unity and $\sigma$ is the standard deviation of the density field
normalized by its mean value (i.e. $\rho/\rho_0$).  When taking the logarithm of the normalized density field this relationship becomes:
\begin{equation}
 \sigma_{s}^2=ln(1+b^2{\cal M}_s^2)
\label{eq:logvar}
\end{equation}
where $s=ln(\rho/\rho_0)$ and $\sigma_{s}$ is the standard deviation of the logarithm of density (not to be confused with $\sigma_{\rho/\rho_0}$).
The relationships above, including the values for $b$, have been empirically derived from MHD and hydrodynamic numerical simulations.
Generally, the value of $b$ depends on the driving of the turbulence in question with $b=1/3$ for solenoidal forcing and $b=1$ for compressive driving
(Nordlund \& Padoan 1999; Federrath, Klessen, \& Schmidt 2008; Federrath et al. 2010). 
Recently, Molina et al. (2012) derived several variants of Equation \ref{eq:logvar}  \textit{analytically} from the MHD shock jump conditions, including the effects 
of strong and weak correlations between magnetic field and 3D density.

A key limitation of these $\sigma^2$-${\cal M}_s$ studies is that the relationships
are derived \textit{only for 3D density}, which is not an observable quantity.  
In this letter, we investigate the applicability of Equations \ref{eq:linvar} and \ref{eq:logvar} for synthetic column density maps, in order to make
these methods more easily applicable to observations. In this case, we define $\zeta=ln(\Sigma/\Sigma_0)$, were $\Sigma$ is the 2D column
density distribution available from the observations. The organization of this letter is as follows:
In section \ref{num} we describe our numerical set up. In Section \ref{res} we discuss the $\sigma^2$-${\cal M}_s$ relationship for
column density. We discuss the results in Section \ref{disc} followed by the conclusions in Section \ref{con}.

\section{Numerical Setup}
\label{num}
We generate a database of 31 3D numerical simulations of isothermal compressible (MHD)
turbulence with resolution $512^3$ and $256^3$. We use the MHD code detailed in
Cho \& Lazarian 2003 and vary the input 
values for the sonic and Alfv\'enic Mach number.
 We briefly outline the major points of the numerical setup. 

\begin{table}
\begin{center}
\caption{Description of the MHD simulations}
\label{tab:models}
\begin{tabular}{ccccc}
\hline\hline
Model & ${\cal M}_{s,rms}$ & ${\cal M}_{A,}$ & Plasma $\beta$ &Resolution  \\
\tableline
1 &8.8 &1.4& 0.05 &$512^3$ \\
2 &7.5 &0.5&0.01 &$512^3$\\
3 &7.3 & 1.5&0.08 &$512^3$ \\
4 &6.1 &0.5&0.01&$512^3$\\
5 &5.8 &1.7&0.17&$512^3$ \\
6 &5.4 &0.5&0.02&$512^3$  \\
7 &3.6 &1.5&0.34&$512^3$ \\
8 &3.7 &0.5&0.04&$512^3$  \\
9 &2.8 &1.7&0.7&$512^3$ \\
10 &2.7 &0.6&0.1&$512^3$ \\
11 &2.1 &1.9&2.4&$512^3$ \\
12 &2.2 &0.7&0.2&$512^3$ \\
13 &0.8 &1.7&9.0&$512^3$   \\
14 &0.8 &0.7&1.5&$512^3$   \\
15 &0.6&1.7&16.1&$512^3$ \\
16 &0.6 &0.7&2.7 &$512^3$  \\
17 &0.4 &1.7&36.1 &$512^3$   \\
18 &0.4 &0.7&6.1 &$512^3$   \\
19&0.7& 2.7&29.7&$256^3$ \\
20&2.2 &3.2&4.1&$256^3$  \\
21&2.0&2.0&2.0&$256^3$ \\
22 &0.7&1.9&14.7&$256^3$  \\
23 &0.9 &1.9&8.9&$256^3$  \\
24 &6.0& 0.5&0.01&$256^3$ \\
25 &0.7 &0.6&1.5&$256^3$  \\
26 &2.0 &0.6&0.2&$256^3$  \\
27 &0.9& 0.7&1.2&$256^3$ \\
28 &1.0 &0.5&0.5&$256^3$  \\
29 &2.7&0.4&0.04&$256^3$  \\
30 &1.0 &0.3&0.2&$256^3$  \\
31 &3.0 &0.3&0.02&$256^3$  \\

\hline\hline
\end{tabular}
\end{center}
\end{table}

The code is a second-order-accurate ENO scheme which solves
the ideal MHD equations in a periodic box.
We drive turbulence solenoidally with energy injected on the large scales.
The magnetic field consists of the uniform background field and a 
fluctuating field: ${\bf B}= {\bf B}_\mathrm{ext} + {\bf b}$. Initially ${\bf b}=0$.
We stress that simulations are scale free
and all units are related to the turnover time and energy 
injection scale.
We divided our models into two groups corresponding to 
sub-Alfv\'enic ($B_\mathrm{ext} \geq 1.0$), 
super-Alfv\'enic ($B_\mathrm{ext} \leq  0.1$) turbulence. Initial values of the Alfv\'enic Mach number for $B_\mathrm{ext} =  0.1$ cases
are $\approx$ 7.0 but decrease as the magnetic field is amplified due to turbulence.
For each group we computed several models with different values of 
gas pressure (see Table \ref{tab:models}) falling into regimes of subsonic and supersonic.  We also include the plasma $\beta$, i.e.  $\beta=P_{thermal}/P_{mag}$,
for each of our simulations in Table \ref{tab:models}.
We create the synthetic column density maps by integrating the 3D density cube along a given line-of-sight (LOS)
parallel or perpendicular to the mean field.
We calculate the average PDFs for three different sight-lines 
in order to take into account LOS effects. 

\section{The Column Density variance- Mach number relation}
\label{res}
\begin{figure}[tbh]
\centering
\includegraphics[scale=.75]{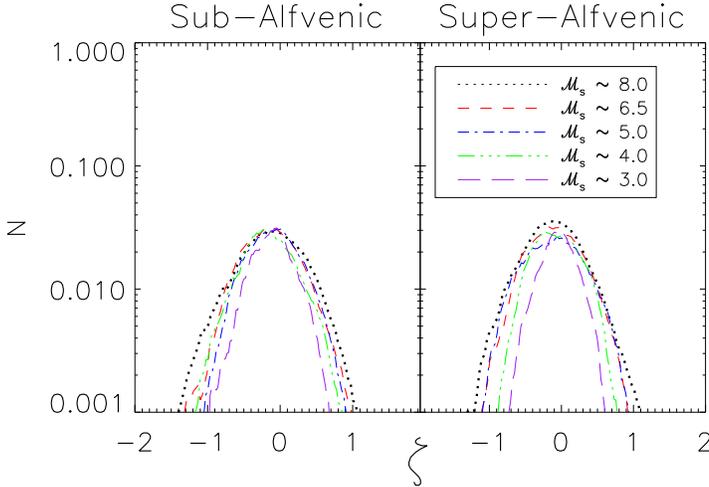}
\caption{ PDFs of $ln(\Sigma/\Sigma_0)$.  
The left panel shows sub-Alfv\'enic models and the right panel shows super-Alfv\'enic models.
All models plotted here have resolution $512^3$.  All models show log-normal PDFs over a range of sonic and Alfv\'enic
Mach number. }
\label{fig:log}
\end{figure}

We calculate the PDFs of the synthetic column density maps 
of all models listed in Table \ref{tab:models} for both the linear distribution (i.e. $\Sigma/\Sigma_0$) and taking the natural logarithm of the maps 
(i.e. $\zeta=ln(\Sigma/\Sigma_0)$).
Examples of the PDFs of the logarithmic column density ($\zeta$) can be found in Figure \ref{fig:log}.
The left panel shows sub-Alfv\'enic models while the right panel shows super-Alfv\'enic models.
All PDFs shown here are taken from simulations that have resolution $512^3$.  All of our synthetic column densities
 show log-normal PDFs. Visually, it is clear that the higher the sonic Mach number, the larger the PDF width.  Thus, we expect the same
trend found in the case of 3D density (i.e. variance increases with ${\cal M}_s$) to hold for 2D column density distributions.

While it maybe the case that the variance will increase with ${\cal M}_s$ for column density, 
Equation \ref{eq:logvar} is derived for 3D density and will not fit the 2D distribution. We note that Equation \ref{eq:logvar}
is  not the only one of its kind found in the literature.  
For example, Lemaster \& Stone (2008) 
used a three parameter fit for the density mean-${\cal M}_s$ relation while 
Molina et al. (2012) derived a variant of Equation \ref{eq:logvar} from the shock jump conditions, including an additional parameter for plasma $\beta$.
These derivations were again only done for the case of 3D density fields, 
and not for observable column density maps.  

We can expect the 2D column density to have similar behavior to the 3D density field (i.e. will be log-normal) in the limit that the 
integration column be smaller than, or comparable to the correlation
length of the turbulence (Vazquez-Semandeni \& Garcia 2001). 
We therefore consider a form similar to Equation \ref{eq:logvar},
but include a new scaling parameter $A$ along with the parameter $b$. Thus, the relationship for $\zeta=ln(\Sigma/\Sigma_0)$ is:

\begin{equation}
\sigma_{\zeta}^2=A\times ln(1+b^2{\cal M}_s^2)
\label{eq:logvar_cd}
\end{equation}

For a log-normal distribution, the linear and logarithmic variances are related by: $\sigma_{\Sigma/\Sigma_0}^2=exp(\sigma^2_{\zeta})-1$, 
Thus the corresponding relation for the linear variance based on Equation \ref{eq:logvar_cd} is:
\begin{equation}
  \sigma_{\Sigma/\Sigma_0}^2=(b^2{\cal M}_s^2+1)^A-1
\label{eq:linvar_cd}
\end{equation}

\begin{figure*}[tbh]
\centering
\includegraphics[scale=.8]{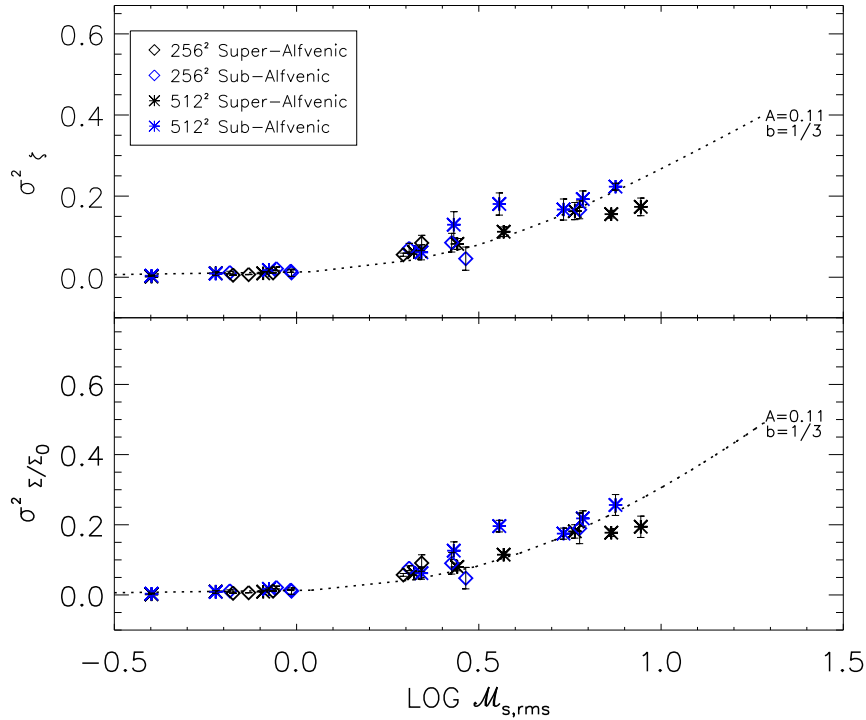}
\caption{The column density variance- ${\cal M}_s$ relationship ($\sigma_{\Sigma/\Sigma_0}^2$ vs ${\cal M}_s$). 
Blue symbols are for sub-Alfv\'enic simulations and black for super-Alfv\'enic cases.
Diamond symbols represent simulations with resolution for $256^3$ while astrix symbols are for cases with resolution $512^3$.
Error bars are created by taking the standard deviation of values between different sight lines. The black dotted lines represent 
 Equation \ref{eq:logvar_cd} and \ref{eq:linvar_cd} with $A=0.11$ and  $b=1/3$ for the top and bottom plots, respectively. }
\label{fig:logpdf}
\end{figure*}

The density variance- ${\cal M}_s$ relationship of Equation \ref{eq:logvar} has only one fit parameter, and here we have chosen a two-parameter fit with $b$, and $A$.
However, due to the similarity between column density and density statistics, we can take the value of $b$ to be the value expected for solenoidal driving, i.e. $b=1/3$.
In this case, we find the best fit $A$ to be $A=0.11$.
We plot the column density variance- ${\cal M}_s$ relationship in Figure \ref{fig:logpdf} for the logarithmic distribution ($\zeta$, top) and the linear distribution (bottom).
Blue symbols are for sub-Alfv\'enic simulations and black for super-Alfv\'enic cases.
Diamond symbols represent simulations with resolution for $256^3$ while asterisk symbols are for cases with resolution $512^3$.
Error bars are created by taking the standard deviation of values between different sight lines. The black dotted line represents 
Equation \ref{eq:logvar_cd} and \ref{eq:linvar_cd} with $A=0.11$ and  $b=1/3$ for the top and bottom plots, respectively. 

The column density variance- ${\cal M}_s$ relationship is narrower for low values of sonic Mach number and slightly wider at higher sonic Mach numbers.  This is due
to observing at filamentary supersonic turbulence along different sight-lines. 
Examination of Figure \ref{fig:logpdf} shows that the column density variance- ${\cal M}_s$  has a slight dependency
on the magnetic field within the error bars, however it is not a overwhelming trend.  
This was also reported in other studies of PDF moments (Kowal, Lazarian, \& Beresnyak 2007).
There is a general increase of the variance with sub-Alfv\'enic turbulence, which can also be visibly seen in Figure \ref{fig:log}.
This is probably related to the effects studied in Molina et al. (2012), who found that the $\sigma^2_{\rho/\rho_0}$-${\cal M}_s$ relationship has a dependency on the 
plasma $\beta$ in the super-Alfv\'enic MHD regime and that this relation breaks down in the sub-Alfv\'enic regime due to increased anisotropy.  
In the case of column density, an additional parameter that will affect the observed anisotropy is the line-of-sight chosen. 
Both the level of the observed anisotropy and the correlations
between $\rho$ and B change as the line-of-sight changes (Burkhart et al. 2009).  
In this case, it is beyond the scope of this letter to examine how the observed line-of-sight changes the $\sigma^2_{\rho/\rho_0}$-${\cal M}_s$ 
relationship of Molina et al. (2012) when translated to 2D column density.  

We find that there is no substantial difference between models with $256^3$ and $512^3$ resolution at the currently studied sonic Mach numbers.
This is not surprising in the context of other works that studied the density variance.  Price \& Federrath (2010) found that
the PDFs for both grid and SPH codes at $256^3$ and $512^3$ converge. 
However, PFB11 found that, at-${\cal M}_s > 5$  the linear variance was affected by numerical resolution, while logarithmic variance is independent of numerical
resolution. At ${\cal M}_s >5$ our simulations become more spread about the expected trend, however it is not clear if this is due to numerical issues or
line-of-sight effects.  Future column density studies extending to higher values of ${\cal M}_s$ are needed to confirm this.

\subsection{Observations}
With the wealth of integrated intensity and dust extinction maps, why are there so few \textit{observational} studies investigating the relationship between $\sigma^2$-${\cal M}_s$?
The answer, in part, is because it is not straightforward to calculate $\sigma^2_{\rho/\rho_0}$ from $\sigma^2_{\Sigma/\Sigma_0}$, which is required for the
application of Equation \ref{eq:logvar}.
Brunt (2010) and PFB11 investigated the relationship of  $\sigma_{\rho/\rho_0}^2$-${\cal M}_s$ for Taurus and calculated the column density variance
using a combination of $^{13}$CO and dust extinction maps. They found ${\cal M}_s \approx 17.6 \pm 1.8$ for Taurus using velocity dispersions. However, 
this value was calculated incorrectly and an erratum is in preparation.  The corrected value of ${\cal M}_s$ for Taurus is $\approx 10$ 
(private communication with C. Brunt, also see Kainulainen \& Tan 2012).
Brunt (2010) finds $\sigma_{\Sigma/\Sigma_0,Av}^2=0.7 \pm 0.04$ (with the error coming from the noise variance) for the Taurus dust extinction maps taken from Froebrich et al. (2007).
Although Brunt (2010) used the $^{13}$CO and the dust extinction maps to obtain the underlying column density variance, we feel that the use of the dust
maps is more reliable due to a number of effects (i.e.$^{13}$CO is more susceptible to
excitation effects, optical depth effects, and has a small dynamic range of density), which we detail more in Section \ref{disc}. 
In addition to the Taurus molecular cloud, Padoan et al. 1997 investigated the $\sigma^2$-${\cal M}_s$ in the cloud IC 5146 using dust extinction observations by Lada et al. (1994).
They found ${\cal M}_s \approx 10$ and  $\sigma_{\Sigma/\Sigma_0,Av}^2=0.49\pm 0.01$, which they then used to convert to a 3D density variance.
The values of $b$ are conservatively constrained from $0.3-0.8$ for these molecular clouds (Brunt 2010) with $b=0.48-0.5$ being the most agreed upon value (Padoan  et al. 1997; Brunt 2010). 
For Taurus, the value of $b$ is still $\approx$ 0.5 despite the change in the sonic Mach number (see Kainulainen \& Tan 2012).

Because we are interested in investigating how the $\sigma_{\Sigma/\Sigma_0}^2$-${\cal M}_s$ relationship can be directly applied to the observations, we 
must consider the effects of instrument noise and telescope smoothing on our synthetic column density maps.
We apply Gaussian white noise with mean signal-to-noise=100 and smoothing with a Gaussian beam to the models listed in Table \ref{tab:models}.
We choose a beam size of 4.6 arcminutes, with an assumed distance to our 'cloud' of 140pc and a cloud size of 24pc.

\begin{figure*}[tbh]
\centering
\includegraphics[scale=.8]{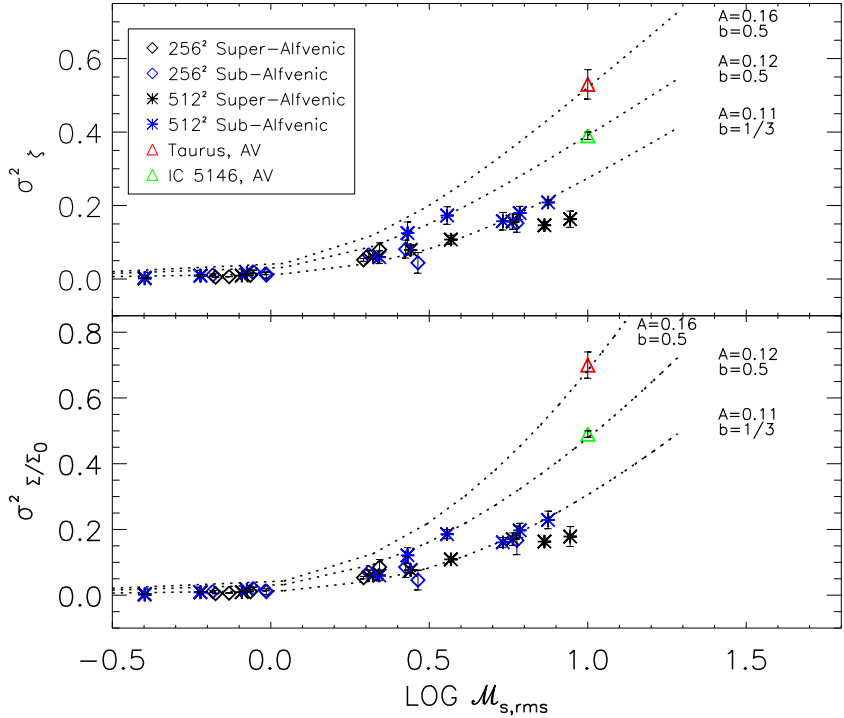}
\caption{Similar to Figure \ref{fig:logpdf} only now the synthetic column density maps include smoothing and Gaussian noise.  
The trend is unaffected at lower Mach number regimes, however at high sonic Mach number (${\cal M}_s > 7.0$) the values for $\sigma^2$
show slightly more of a decrease in variance due to shocks being smoothed over. 
The  Taurus (red triangle; Brunt 2010) and IC 5146 (green triangle; Padoan et al. 1997) clouds are included along with the
fits from Equations \ref{eq:logvar_cd} and \ref{eq:linvar_cd}.}
\label{fig:logpdf_smoo}
\end{figure*}

We plot the column density variance derived from the dust extinction maps for Taurus and IC 5146 and our models from Table \ref{tab:models}, which now include 4.6 arcminute smoothing
and Gaussian noise, in Figure \ref{fig:logpdf_smoo}, in order to test our 
$\sigma^2$-${\cal M}_s$ relationship given in Equation \ref{eq:logvar_cd} and \ref{eq:linvar_cd} on the observations.
The linear column density variance- ${\cal M}_s$ relationship is shown in the bottom panel
and the logarithmic distribution ($\zeta$) in the top panel.
Blue symbols are for sub-Alfv\'enic simulations and black for super-Alfv\'enic simulations.
Diamond symbols represent simulations with resolution for $256^3$ while asterisk symbols are for cases with resolution $512^3$.
The black lines represent Equation \ref{eq:logvar_cd} and \ref{eq:linvar_cd} (for the top and bottom plots, respectively) 
with $b=1/3$ and $A=0.11$ for the isothermal simulations
and $b=0.5$ and $A=0.16$ and  $b=0.5$ and $A=0.12$ for Taurus and IC 5146, respectively.
The green and red triangles are the $\sigma_{\Sigma/\Sigma_0}^2$-${\cal M}_s$ values from the dust extinction maps for the IC 5146 and Taurus molecular clouds, as given by 
Padoan et al. (1997) and Brunt (2010), respectively.

Comparison of Figure \ref{fig:logpdf} (where no smoothing is included)
with Figure \ref{fig:logpdf_smoo} shows that the inclusion of smoothing and noise, at least at the level added here, does not severely affect
the applicability of Equations \ref{eq:logvar_cd} and \ref{eq:linvar_cd}. In general, smoothing decreases the value
of  $\sigma^2$, causing one to underestimate ${\cal M}_s$.

The values for the Taurus and IC 5146 clouds lie very close to what is expected from Equations \ref{eq:logvar_cd} and \ref{eq:linvar_cd} 
for $b=0.5$ and $A=0.16$ and   $b=0.5$ and $A=0.12$, respectively.  Although Equations \ref{eq:logvar_cd} and \ref{eq:linvar_cd} 
fit well,  it is clear that the solenoidal
simulations  do not fit well with the molecular cloud data.  This was also the case in PFB11, who 
attributed the discrepancy between their simulations and observations to the lack of self-gravity and/or use of purely solenoidal forcing\footnote{Molecular clouds most likely
have a mixed solenoidal and compressive forcing environment, which should yield higher values of $b$, see Federrath et al. (2010)}. 
For PFB11 the assumptions involved to convert the 2D to 3D variance and the use of CO in the variance calculation may also effect the relation.
The fact that the dust extinction observations are able to fit Equations \ref{eq:logvar_cd} and \ref{eq:linvar_cd} using the values of $b$ obtained
from previous studies opens up the prospects of using the 2D column density variance in addition to the assumed 3D variance.  
Additionally, the parameter $A$ may have dependencies on other physics, which are discussed below.

\section{Discussion}
\label{disc}

In this letter we found a relation between the variance of column density
and ${\cal M}_s$. The 3D density variance relationship of Equation \ref{eq:logvar} was modified to obtain the $\sigma^2$-${\cal M}_s$ relationship for column density. 
This was motivated by the fact that $\sigma_{\rho/\rho_0}^2$ is not available from observations.  
Ideally, both $\sigma_{\rho/\rho_0}^2$, as determined by the method outlined in Brunt, Federrath, Price (2010), and the direct  $\sigma_{\Sigma/\Sigma_0}^2$-${\cal M}_s$ relationship outlined
here should be used together to provide more confidence in the $\sigma^2$-${\cal M}_s$ relation and help constrain the values of $b$.

Furthermore, the results of this letter  can be used synergistically with
higher order PDF moments suggested in a number of recent papers (see Kowal et al. 2007, Burkhart et al. 2009) to study ${\cal M}_s$. 
The use of different approaches in determining ${\cal M}_s$ from observed column densities including other techniques, e.g. utilizing Tsallis statistics 
(Esquivel \& Lazarian 2010, Tofflemire et al. 2011) and the genus measure (see Burkhart, Lazarian \& Gaensler 2012), can further increase the reliability of the results obtained.

An additional complicating factor for the observations is that obtaining the true column density PDF can be tricky (Goodman, Pineda \& Schnee 2009).
In general, the use of dust extinction maps is more reliable than $^{13}$CO in order to obtain the true column density for a variety of reasons, including a larger dynamic range
(factors of 50 or more, see Goodman, Pineda \& Schnee 2009, for more detailed discussion).
Pineda, Caselli,\& Goodman (2008) showed that the $^{13}$CO derived column density estimates are adversely effected by optical depth as low as A$_v$ $\approx$ 4 mag.
Molecular transitions have a limited range of volume  densities that they can trace, since below a critical density there may not be enough molecules to excite the
transition and at very high densities optical depth effects will mask the true column density. 
For these reasons, we chose to plot the values of $\sigma^2$ for Taurus and IC 5146 taken from the dust extinction maps, rather than the $^{13}$ CO maps.  

Parameters $A$ and $b$ have dependencies on external physics beyond the sonic Mach number.
$b$  depends on the
driving of the turbulence and it is possible that $A$ may follow the same behavior.  In order to test this, we 
take values from Table 3 of Federrath et al. 2010, with $\sigma_{\Sigma/\Sigma_0, sol}^2$=0.46 and $\sigma_{\Sigma/\Sigma_0, com}^2$=1.51 and 
${\cal M}_s$=5.5. We note that these values are taken from simulations that have purely solenoidal or purely compressive forcing,
when in realty molecular clouds most likely have turbulence driven with mixed forcing.
 We find that applying these values to Equation \ref{eq:logvar_cd} yields $A_{sol}=0.14$ and $A_{com}=0.6$. However,
when using a simple scaling between $\sigma_{\Sigma/\Sigma_0}$ and $\sigma_{\rho/\rho_0}$ (given in Table 1 of Federrath et al. 2010), 
we find that $A_{sol}=0.12$ and $A_{com}=0.25$.  
Both methods show that the parameter $A$ has a significant dependency on the type of driving, however there is
large variation in the compressive driven values of $A$ based on these tables, which may be due to LOS effects or
insufficient statistics. Additionally, some dependency on observational effects such as beam smoothing and noise was seen in Figure \ref{fig:logpdf_smoo} and
future studies will determine what the effects of optical depth my have on these parameters. 

The fact of the matter is that the variance of the density or column density
distribution does not only depend on the sonic Mach number, but a host of other contributing factors.  This is strong motivation for the use of \textit{multiple} tools and techniques
to obtain information on the turbulent state of the gas.  A collaborative use of PDF methods that include the 3D and 2D variance
as well as higher order moments of the linear distribution will yield the most accurate picture.

\section{Conclusions}
\label{con}
We investigate the utility of a 2D column density $\sigma_{\Sigma/\Sigma_0}^2$-${\cal M}_s$ trend which can be used in addition to the existing 3D density
$\sigma_{\rho/\rho_0}^2$-${\cal M}_s$ trend. We find that:

\begin{itemize}
\item Equations \ref{eq:logvar_cd} and \ref{eq:linvar_cd} (for logarithmic
and linear variance, respectively) empirically fit
synthetic column density maps well, even when including observational effects such as smoothing and noise,
with $b=1/3$ (as appropriate for the solenoidal forcing used in this study, see Federrath et al. 2008) and $A=0.11$.
\item For the dust extinction maps of the Taurus and IC 5146 molecular clouds, we find $A=0.16$ and $A=0.12$, respectively, when using
 $b=0.5$ (as given in the literature) for the $\sigma_{\Sigma/\Sigma_0}^2$-${\cal M}_s$ relation.
\end{itemize}

\acknowledgments
We thank the referee, Dr. Christoph Federrath, for his valuable comments.
B.B. acknowledges support from the
NSF Graduate Research Fellowship and the NASA Wisconsin Space Grant.
A.L. acknowledges NASA grant NNX11AD32G, the Vilas Award and the Center for Magnetic Self-Organization in Astrophysical and Laboratory Plasmas for financial support.

\end{document}